\documentclass[conference]{IEEEtran}
\IEEEoverridecommandlockouts
\usepackage{cite}
\usepackage{amsmath,amssymb,amsfonts}
\usepackage{algorithmic}
\usepackage{graphicx}
\usepackage{textcomp}
\usepackage{xcolor}
\usepackage{units}
\usepackage{multirow}
\usepackage{colortbl}
\usepackage{soul}
\usepackage{booktabs}
\usepackage[left=0.625in, right=0.625in,top=0.75in,bottom=1in]{geometry}

\usepackage{amsthm}
\usepackage{thmtools, thm-restate}
\usepackage{algorithm}

\ifodd 1

\newcommand{\com}[1]{{\color{red}\textbf{Comment}: #1}}
\newcommand{\comtoo}[1]{{\color{purple}\textbf{Comment}: #1}}
\newcommand{\resp}[1]{{\color{cyan}\textbf{Response}: #1}} 
\else

\newcommand{\com}[1]{}
\newcommand{\comtoo}[1]{}
\newcommand{\resp}[1]{}
\fi

\def\BibTeX{{\rm B\kern-.05em{\sc i\kern-.025em b}\kern-.08em
    T\kern-.1667em\lower.7ex\hbox{E}\kern-.125emX}}
\begin{document}

\title{AI-Driven Spectrum Occupancy Prediction Using Real-World Spectrum Measurements}

\author{
\IEEEauthorblockN{Jiayu Mao\IEEEauthorrefmark{2}, 
Ruoyu Sun\IEEEauthorrefmark{1},
Mark Poletti\IEEEauthorrefmark{1},
Rahil Gandotra\IEEEauthorrefmark{1},
Hao Guo\IEEEauthorrefmark{1},
Aylin Yener\IEEEauthorrefmark{2}}

\IEEEauthorblockA{\IEEEauthorrefmark{1}Department of Wireless Technologies, CableLabs, Louisville, Colorado, USA}
\IEEEauthorblockA{\IEEEauthorrefmark{2}Department of Electrical and Computer Engineering, The Ohio State University, Columbus, Ohio, USA}
Email: mao.518@osu.edu, \{r.sun, m.poletti, r.gandotra, h.guo\}@cablelabs.com,  yener@ece.osu.edu

\thanks{This paper is under review at an IEEE conference.}
}

\newgeometry{left=0.625in, right=0.625in,top=0.75in,bottom=1in}
\maketitle

\begin{abstract}
Spectrum occupancy prediction is a critical enabler for real-time and proactive dynamic spectrum sharing (DSS), as it can provide short-term channel availability information to support more efficient spectrum access decisions in wireless communication systems. 
Instead of relying on open-source datasets or simulated data, commonly used in the literature, this paper investigates short-horizon spectrum occupancy prediction using mid-band, 24×7 real-world spectrum measurement data collected in the United States.
We construct a multi-band channel occupancy dataset through analyzing 61 days of empirical data and formulate a next-minute channel occupancy prediction task across all frequency channels. 
This study focuses on AI-driven prediction methods, including Random Forest, Extreme Gradient Boosting (XGBoost), and a Long Short-Term Memory (LSTM) network, and compares their performance against a conventional Markov chain-based statistical baseline. 
Numerical results show that learning-based methods outperform the statistical baseline on dynamic channels, particularly under fixed false-alarm constraints. 
These results demonstrate the effectiveness of AI-driven spectrum occupancy prediction, indicating that lightweight learning models can effectively support future deployment-oriented DSS systems.

\end{abstract}

\section{Introduction}
\label{sec:intro}
The radio spectrum is a limited natural resource.
As wireless services continue to expand, such as spanning cellular, Wi-Fi, satellite, and emerging sensing-enabled applications, the demand for reliable spectrum access is growing, while interference constraints and incumbent allocations limit opportunities for reuse.
In the United States, spectrum management is shared between the Federal Communications Commission (FCC), which regulates non-federal spectrum use, and the National Telecommunications and Information Administration (NTIA), which manages federal spectrum use, highlighting the need for careful coordination to enable efficient use of shared bands~\cite{fcc}.
Practical measurements indicate that spectrum utilization is highly time-varying and band-dependent; therefore, dynamic spectrum sharing (DSS) is essential to improve spectrum utilization efficiency and reduce avoidable spectrum conflicts~\cite{sagduyu2024joint}.
In particular, DSS can benefit from short-horizon prediction of channel availability, which relies on active spectrum sensing.
This motivation is further strengthened by the growing emphasis on wide-area spectrum sensing and scanning capabilities in 6G vision and standardization activities, as reflected in ongoing 3GPP efforts (e.g., TR 22.870, S1-254270)~\cite{3gpp_tr_22_870,3gpp_s1_254270}.

Spectrum occupancy prediction has emerged as a key enabler for proactive spectrum management in 6G era, wherein future spectrum availability is predicted from historical spectrum utilization data using statistical and machine learning (ML) approaches~\cite{wang2024deep}.
Although spectrum sensing~\cite{xiong2013spectrum} provides instantaneous channel state information, spectrum prediction leverages historical sensing data and their post-processing information to infer near-future spectrum occupancy and availability.
To facilitate spectrum prediction, channel occupancy~\cite{raouf2025spectrum,yin2012mining} is commonly used to represent spectrum usage, capturing the temporal behavior of occupied and idle channel states and enabling the development of both statistical and learning-based prediction methods.
This representation enables spectrum predictive capabilities that simplify channel selection, reduce unnecessary sensing operations, and mitigate potential congestion.
Moreover, frequent spectrum reallocation and coexistence procedures can introduce non-trivial signaling overhead and access latency, particularly in highly-dynamic environments; short-horizon spectrum occupancy prediction therefore supports faster and more efficient spectrum-access decisions.

Existing studies on spectrum prediction can generally be grouped into model-driven and data-driven approaches. 
Model-driven methods focus on statistical analysis that are known for simplicity and low computational cost.
For example,~\cite{chen2014predicting} proposes a non-stationary hidden Markov model to capture time-varying channel occupancy.
Reference~\cite{wen2008autoregressive} presents an auto-regressive channel prediction model for cognitive radio systems.
However, these methods often struggle to deal with large datasets and complicated non-linear spectrum usage patterns. 
With the increasing availability of long-term spectrum measurement data, conventional statistical methods continue to face challenges in exploiting such rich information. 
In contrast, data-driven approaches based on machine learning have received significant attention recently.
References~\cite{zhang2023gcrnn} and~\cite{li2025multi} propose attention-based graph neural network models that exploit correlations across multiple frequency bands for spectrum prediction.
In addition,~\cite{alahakoon2025lstm} and \cite{xu2024spectrum} employ long short term memory (LSTM)-based models to predict spectrum occupancy by exploiting temporal correlations in historical sensing data.
Similarly, LSTM-based prediction is compared with auto-regressive models in~\cite{mosavat2021prediction}.
Reference~\cite{ji2025generative} proposes a generative broad learning framework.
More recently, transformer-based methods have been applied to spectrum prediction in~\cite{liu2025spectrumllm} and \cite{li2025novel}, capturing long-range dependencies and complex correlations at the cost of increased model complexity and computational overhead.
Beyond spectrum prediction, learning-based techniques have also been integrated into spectrum management frameworks, where predictive information is leveraged to support proactive resource allocation and channel selection in dynamic wireless environments~\cite{zhou2025spectrumfm}.
While learning-based methods improve prediction accuracy and robustness, they often incur higher computational complexity, motivating interest in lightweight models suitable for practical deployment.

In contrast to the majority of existing works that rely on simulation data or open-source datasets, this paper is based on real spectrum measurement data collected over an extended period. Using these measurements, a multi-channel channel-occupancy dataset is constructed and a short-horizon spectrum occupancy prediction task that targets next-minute availability across frequency bins is formulated. 
This study focuses on AI-driven spectrum prediction and evaluates representative machine learning models, including Random Forest, XGBoost, and a LSTM network, while using a first-order Markov chain as a statistical baseline for comparison under a unified problem formulation and evaluation framework.
Beyond aggregate accuracy, we examine the impact of channel dynamics, class imbalance, and history length on prediction performance, and assess detection capability under fixed false-alarm constraints. The results provide practical insights into when learning-based methods offer tangible benefits over the statistical baseline and highlight the effectiveness of lightweight models such as Random Forest for deployment-oriented dynamic spectrum sharing systems.

The remainder of the paper is organized as follows: Section II introduces the problem formulation, Section III describes the machine learning methods, Section IV presents numerical results, and Section V concludes the paper.

\section{Problem Formulation} 
\label{sec: sysmodel}
This section first describes the data collection process using the spectrum analyzer and outlines the subsequent post-processing steps used to obtain channel occupancy information. 
Channel occupancy is then formally defined as a processed representation of spectrum utilization over time and frequency~\cite{yin2012mining}. 
Based on this representation, the machine learning task of predicting short-term channel availability is formulated. 
Specifically, the construction of the dataset, the input features, and the target variables to be predicted are described. 
The detailed descriptions of the learning algorithms employed for this task are provided in a later section.

\subsection{Channel Occupancy} 
\label{subsec: CO}
\begin{figure}[t]
    \centering
    \includegraphics[width=0.9\linewidth]{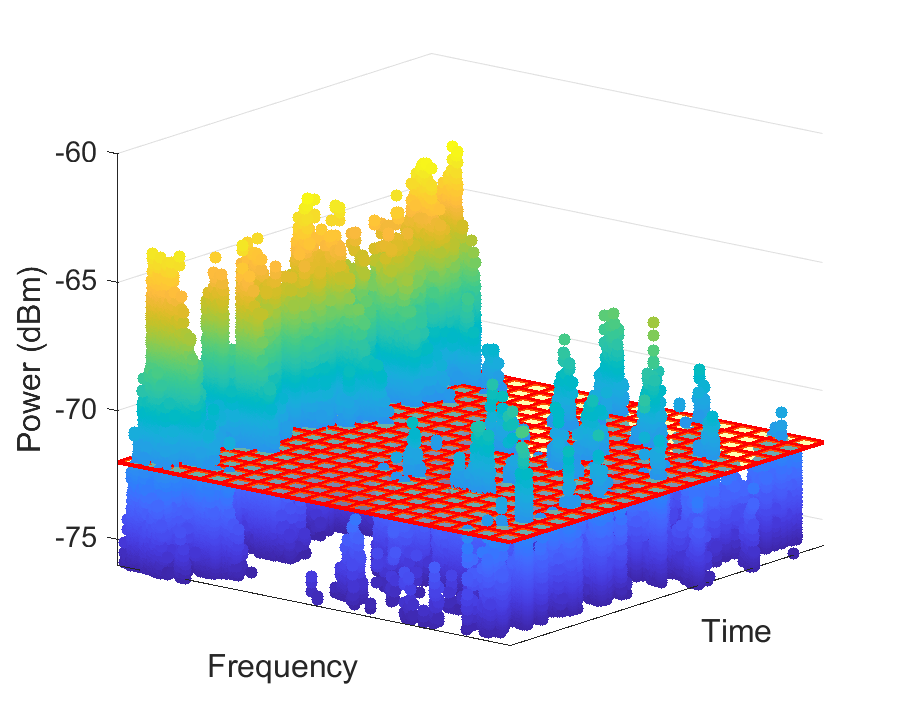}
    \caption{3D raw power spectrum data.}
    \label{fig:raw3D}
\end{figure}

In our prior works~\cite{mark2024scte,raouf2025spectrum,rahilvtc,guovtc}, an automated spectrum utilization monitoring system is developed that collects spectrum scanning data from multiple remote sites, where antennas and spectrum scanners capture signals, and local computers record and compress the collected data. 
The data is automatically transferred to a central location via FTP, where it is processed, stored in a database, and visualized through a web-based dashboard. 
To quantify how the monitored spectrum is being used, the concept of channel occupancy is introduced, which provides a binary measurement of whether a channel is used during a given time interval.
Specifically, the channel is considered \emph{occupied} if at least one received data sample within the interval exceeds a predefined power threshold, and \emph{idle} otherwise. 
Formally, the channel occupancy $CO_{f,t}$ at frequency bin $f$ and time index $t$ is defined as
\begin{equation}
CO_{f,t} =
\begin{cases}
0, & \text{if no sample exceeds the threshold,} \\
1, & \text{if at least one sample exceeds the threshold}.
\end{cases}
\end{equation}
This binary representation provides a compact and tractable description of spectrum utilization over time and frequency, which will serve as the foundation for the subsequent machine learning problem formulation.
Fig.~\ref{fig:raw3D} illustrates an example of the raw power spectrum measurements collected over frequency and time. 
The vertical axis represents the received power in dBm, while the red horizontal plane denotes the predefined power threshold. 
Whenever the measured power at a given frequency-time point exceeds this threshold, the corresponding channel occupancy is set to $1$; otherwise, it is set to $0$. 

\begin{figure*}[t]
    \centering
    \includegraphics[width=0.9\linewidth]{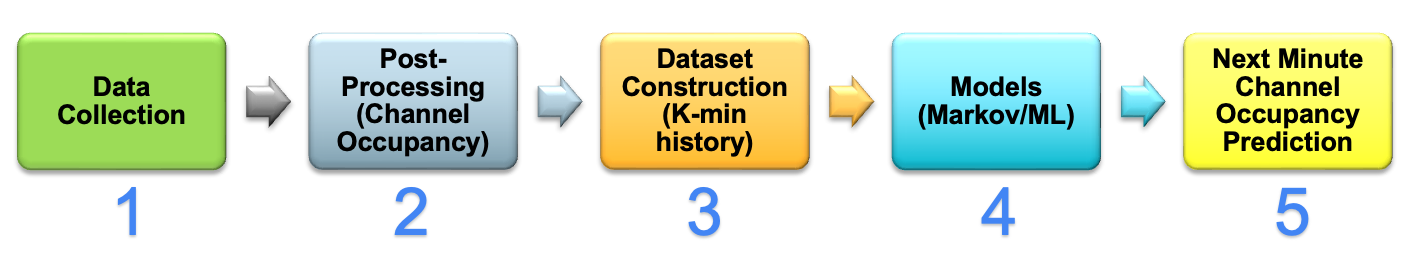}
    \caption{Spectrum utilization prediction pipeline.}
    \label{fig:pipeline}
\end{figure*}
\subsection{Machine Learning Task} 
\label{subsec: MLprob}
This study aims to predict the short-term spectrum availability by formulating channel occupancy prediction as a supervised machine learning problem. Specifically, given the past $K$ minutes of channel occupancy states across the monitored spectrum, the objective is to predict the occupancy in the next minute for all frequency bins. 
This next-minute prediction provides a basis for enabling more efficient and dynamic spectrum sharing strategies, as it aligns with the decision timescales of practical spectrum access and scheduling mechanisms. 
Formally, the task is to learn a mapping
\begin{equation}
f: \{CO_{f,t-K}, \ldots, CO_{f,t-1}\}_{f=1}^{F} \mapsto \{CO_{f,t}\}_{f=1}^{F},
\end{equation}
where $CO_{f,t} \in \{0,1\}$ denotes the occupancy state of frequency bin $f$ at time $t$, and $F$ is the total number of frequency bins.

\vspace{0.5em}
\noindent \textbf{Dataset Construction:}
Spectrum measurements at the location in Louisville, CO over a two-month period (May–June 2025) in the $f_1$--$f_2$~GHz band are used as the historical data. The frequency span was divided into 5~MHz bins, resulting in $F=90$ frequency bins. After applying the post-processing described in the previous subsection, the raw power measurements were converted into binary channel occupancy sequences. The complete dataset spans 61 days, each containing $24 \times 60 = 1440$ one-minute intervals, leading to a total of 87,840 time steps.

To construct the learning dataset, we first concatenate the time series of occupancy states into a single chronological sequence. A sliding-window procedure is then applied: for each time index $t$, the input feature vector $X_t$ is formed by concatenating the occupancy states from the past $K$ minutes across all frequency bins, producing an input of dimension $K \times F$. The corresponding label $y_t$ consists of the occupancy states of all $F$ bins at time $t$. Repeating this procedure across the entire time horizon yields the training pairs $\{(X_t, y_t)\}_{t=1}^{N}$, where $N$ is the number of available samples.

\noindent \textbf{Processing Pipeline:}
The overall procedure is summarized in Fig.~\ref{fig:pipeline}. The workflow begins with raw spectrum data collection, followed by post-processing to obtain binary channel occupancy. These sequences are then used to construct the supervised learning dataset via the sliding-window approach. Finally, the dataset is fed into baseline and machine learning models (e.g., Markov chain, Random Forest, etc), which produce the next-minute channel occupancy predictions across all frequency bins. This pipeline enables systematic evaluation of different prediction models under a consistent dataset and task definition.

\section{Machine Learning Methods} 
\subsection{Random Forest}

Random Forest is adopted as our first tree-based learning method.  
A Random Forest can be viewed as an ensemble of decision trees that collectively form a more stable and accurate classifier than any single tree alone.  
Each decision tree partitions the input space through a sequence of simple rules, for example, threshold comparisons on individual features, and assigns a label at the terminal leaf.  
Although such trees are intuitive to interpret, they are highly sensitive to fluctuations in the training data; small perturbations may lead to very different splits and, therefore, highly variable predictions.  
The Random Forest approach mitigates this sensitivity by training many trees in parallel, each on a different bootstrap sample of the data, and using only a randomly selected subset of features at each split.  
By injecting diversity into the individual trees and aggregating their outputs through majority voting, the resulting ensemble achieves substantially improved robustness and generalization performance.

The Random Forest method is found to be particularly useful because it does not impose strong parametric assumptions and naturally accommodates high–dimensional inputs.  
The randomization mechanisms embedded in the algorithm, including sampling training points and selecting feature subsets, encourage
each tree to capture different local patterns in the data.  
When combined through the ensemble, these varied perspectives yield a classifier that can represent nonlinear relationships and generalize well across a broad range of supervised learning tasks.

To apply Random Forest to the channel occupancy prediction problem, a sliding-window representation is leveraged, as introduced in
Section~\ref{subsec: MLprob}.  
Each example consists of the binary occupancy states from the previous $K$ minutes across all $F$ frequency bins, forming a feature vector of size $K \times F$.   
The prediction target is a binary vector
of length $F$, indicating which bins are occupied in the next minute.

A multi-output formulation is adopted because the task requires predicting multiple binary outputs simultaneously.
In particular, we train $F$ independent Random Forest classifiers in parallel, one for each frequency bin.  
All classifiers share the same feature matrix but learn separate decision rules tailored to their respective bins.  
At inference time, each forest aggregates the predictions from its constituent trees, producing a bin-level occupancy decision independently for each frequency bin. 
The ensemble also provides per-bin probability estimates, which we later use in detection-oriented evaluations, such as computing the probability of detection at fixed false-alarm rates.

In a nutshell, the Random Forest approach offers a flexible, nonparametric, and data-driven framework for short-term channel occupancy prediction, allowing us to capture complex spectral usage patterns without requiring modeling assumptions.

\subsection{XGBoost}

Extreme Gradient Boosting (XGBoost) is implemented as a second tree-based method.  
XGBoost is a way of combining many simple decision trees into a single strong classifier by training them sequentially rather than independently. 
At a high level, we start from an initial model that makes a crude prediction for every training example. Trees are then added one by one, where each new tree is trained to correct the errors of the current ensemble.  
In each boosting round, we compute the gradient of a chosen loss function with respect to the current
predictions and fit a new tree that approximates this gradient.  
By taking a small step in the opposite direction of the gradient, scaled by a learning rate, the loss is gradually reduced.  
This process can be viewed as performing gradient descent in the space of decision trees, which allows us to build highly nonlinear decision boundaries while
retaining the interpretability and sparsity structure of tree models.

XGBoost is chosen because it implements this boosting strategy in an efficient and numerically stable way.  
The algorithm uses second-order information (both gradients and approximate Hessians) to decide how to split nodes, applies explicit regularization on tree complexity, and employs histogram-based split finding to handle large datasets.  
For binary classification, which is the case in our setting, we use the binary logistic objective.  
This objective treats each tree ensemble prediction as a logit and passes it through a sigmoid function to obtain a probability of class~1.  
The training procedure minimizes the logistic loss between these probabilities and the true binary
labels.  
The main hyperparameters controlled are the number of boosting rounds (the number of trees in the ensemble), the maximum depth of each tree, and the learning rate that scales the contribution of every new tree.  
Deeper trees allow more complex interactions between features, while a smaller learning rate usually requires more trees, but can lead to better generalization.

XGBoost is applied to the channel occupancy prediction task using the same sliding-window representation as in the Random Forest approach.  
The input feature vector for each example consists of the binary occupancy states over the previous $K$ minutes across all $F$ frequency bins, forming a $K \times F$ dimensional vector that captures both temporal and spectral patterns.  
The target output is a binary vector of length $F$ that represents the occupancy of each bin in the next minute.  
Aiming to predict all bins simultaneously, we adopt a multi-output formulation.  
In particular, each XGBoost classifier is trained per frequency bin.  
Each classifier observes the same feature matrix but learns a separate decision function corresponding to its bin, and all models are trained in parallel.

\subsection{LSTM}

We use a Long Short-Term Memory (LSTM) network as a sequence–learning method to explicitly model temporal dependencies in the channel-occupancy data.  
Recurrent neural networks (RNNs) are designed to process ordered sequences by maintaining a hidden state that is updated at each time step, so that the prediction at the end of the sequence can depend on all previous inputs.  
Vanilla RNNs, however, tend to suffer from vanishing and exploding gradients when the sequence length grows, which makes it difficult for them to retain information over longer horizons.
The LSTM architecture addresses this limitation by introducing a gated memory cell~\cite{lstm}.  
At each time step, the cell uses input, forget, and output gates to decide which parts of the new input to incorporate, which parts of the existing memory to retain, and which parts to expose to the next layers. 
These gating mechanisms allow the network to selectively store and discard information over time, so it can represent both short and long-range temporal patterns in a stable way.

We adopt an LSTM that reads the recent history of spectrum activity as a sequence and produces a joint prediction for all frequency bins at the next minute.  
At each time step within the look-back window, the input to the LSTM is a vector of length $F$ that contains the binary occupancy states of all frequency bins at that minute.  
Over $K$ steps, the network observes a sequence of such vectors, which we arrange as a tensor of shape $(K, F)$ for each sample.  The LSTM processes this sequence step by step and maintains an internal hidden state that summarizes the relevant temporal information accumulated so far.  
After the last time step, we take the final hidden state as a compact representation of the $K$-minute history and feed it into a fully connected layer that outputs a vector of length $F$.  
Each component of this output corresponds to a logit for one frequency bin, which we interpret as the evidence for that bin being occupied or idle in the next minute.

The problem is formulated as a multi-label binary classification, in which each training example is associated with a binary target vector of length $F$ that indicates the true occupancy of all bins at the prediction time.  
The LSTM-based model learns to map the sequence of past occupancy patterns to this target vector by minimizing a standard sigmoid-based classification loss over all bins.  
This formulation enables the network to exploit correlations not only over time but also implicitly across frequency, because all bins share a common recurrent representation of the historical observations.
By design, the LSTM is able to capture recurring usage patterns, periodic behavior, and bursty transitions that arise in practical spectrum utilization traces, providing a complementary view to the tree-based methods considered in our study.

\section{Numerical Results} \label{sec: exp}

The effectiveness of the proposed distributed client selection approach is evaluated through numerical experiments.

\subsection{Training Setup}

We use a common data split and feature representation for all three learning-based methods (Random Forest, XGBoost, and LSTM).  
We take approximately 1.5~months of channel-occupancy measurements as the training dataset and reserve the remaining disjoint 0.5~month as the test dataset, so that all models are evaluated on future data that do not overlap
with the training period.  
For the tree-based methods, we construct the input feature matrix by flattening the sliding window described in Section~\ref{subsec: MLprob}: each example contains the binary occupancy states over the previous $K$ minutes across all $F$ frequency bins, yielding a feature vector of dimension $K \times F$.  
The corresponding label is a binary vector of length $F$ that records the occupancy of all bins in the next minute, and we train a multi-output classifier that learns one decision function per bin from this shared feature matrix.
For the LSTM model, we keep the same windowed representation but feed it to the network as a length-$K$ sequence of $F$-dimensional vectors, allowing the recurrent architecture to process the temporal evolution of occupancy explicitly. 
After training, all methods output per-bin occupancy decisions as well as probability scores for the occupied state, which we subsequently use to compute average test accuracy, balanced accuracy, and probability of detection at fixed false-alarm rates, focusing on dynamically varying bins identified by transition-rate and class-imbalance criteria.

\subsection{Markov-Chain Baseline}

In addition to learning-based approaches, we include a simple first-order Markov-chain baseline that does not rely on model training in the machine-learning sense.  
For each frequency bin, we view the occupancy process as a two-state Markov chain with states ``idle'' (0) and ``occupied'' (1) and estimate the transition probabilities $P_{0 \to 0}$, $P_{0 \to 1}$, $P_{1 \to 0}$, and $P_{1 \to 1}$ directly from the training labels using smoothed empirical counts.  
During testing, given the previous occupancy state, the model produces a probabilistic one-step-ahead prediction of occupancy in the next minute, for example, $P_{0 \to 1}$ or $P_{1 \to 1}$.
This predicted probability is then thresholded to obtain a hard 0/1 decision, using a fixed threshold of 0.5 for accuracy evaluation.
This Markov model serves as a lightweight, statistical baseline that exploits first-order temporal dependence, against which we compare the gains offered by the more expressive machine-learning methods.

\begin{table}[t]
    \centering
    \caption{Main comparison with fixed history length $K=10$.}
    \label{tab:mainK10}
    \setlength{\tabcolsep}{4pt} 
    \begin{tabular}{lcccc}
        \toprule
        \textbf{Method} &
        \shortstack{\textbf{Average}\\\textbf{accuracy}} &
        \shortstack{\textbf{Balanced}\\\textbf{accuracy}} &
        \shortstack{$\boldsymbol{P_\mathrm{d}}$@\textbf{1\%}\\$\boldsymbol{P_\mathrm{fa}}$} &
        \shortstack{$\boldsymbol{P_\mathrm{d}}$@\textbf{5\%}\\$\boldsymbol{P_\mathrm{fa}}$} \\
        \midrule
        Markov (baseline) & 83.76\% & 68.23\% & 25.05\% & 50.44\% \\
        Random Forest     & 86.29\% & 68.87\% & 40.17\% & 59.18\% \\
        XGBoost           & 86.41\% & 69.63\% & 40.53\% & 59.71\% \\
        LSTM              & 86.45\% & 69.62\% & 40.52\% & 59.97\% \\
        \bottomrule
    \end{tabular}
\end{table}

\subsection{Main Results}
We report four aggregate metrics in Table~\ref{tab:mainK10}.  
The \emph{average accuracy} is the average test accuracy over all test examples and all frequency bins, i.e., the fraction of correctly predicted
occupancy labels when we flatten the time--frequency grid.  
The \emph{balanced accuracy} (BA) accounts for class imbalance by averaging the true positive rate (TPR) and true negative rate (TNR) per bin, and then averaging over bins as follows
\begin{equation}
    \mathrm{BA}_f = \tfrac{1}{2}\bigl(\mathrm{TPR}_f + \mathrm{TNR}_f\bigr),
\end{equation}
so that both occupied and idle states contribute equally, even when one state is rare.

The last two columns summarize detection performance at fixed false-alarm levels. 
Here, the detection probability $P_{\mathrm{d}}$ denotes the true positive rate, i.e., the probability of correctly declaring a channel as occupied when it is truly occupied, while the false-alarm probability $P_{\mathrm{fa}}$ denotes the probability of incorrectly classifying an idle channel as occupied. 
For each method, a common decision threshold $\tau$ is applied to the model output scores $s \in [0,1]$, where a channel is predicted as occupied if its score exceeds $\tau$. 
The metric $P_{\mathrm{d}}@1\%P_{\mathrm{fa}}$ is obtained by selecting a threshold $\tau$ such that the overall false-alarm probability satisfies $P_{\mathrm{fa}}(\tau) \approx 1\%$, and then evaluating the detection probability $P_{\mathrm{d}}(s\geq\tau)$. 
Similarly, $P_{\mathrm{d}}@5\%P_{\mathrm{fa}}$ corresponds to a $5\%$ false-alarm operating point. 
In contrast, accuracy-based metrics use a fixed threshold $\tau=0.5$ and do not explicitly constrain the false-alarm rate.

Observed from Table~\ref{tab:mainK10}, all three learning-based methods consistently outperform the first-order Markov baseline across all metrics.  
The gain is most obvious in the detection-oriented measures: at $1\%$ false-alarm rate, the machine learning models significantly increase the probability of detection compared to the Markov chain (about $40\%$ versus $25\%$), while also improving average and balanced accuracy by several percentage points. 
This improvement arises because learning-based models generate predicted occupancy scores that are better separated between occupied and idle states, enabling higher detection probability under strict false-alarm constraints.
Among the learning-based approaches, Random Forest, XGBoost, and LSTM achieve very similar performance, with only marginal differences at the third decimal place.  
Given this similarity in accuracy and $P_\mathrm{d}$, the Random Forest model emerges as an attractive choice, offering competitive performance with relatively simple training and implementation. In the rest of the analysis, we focus on Random Forest due to its simplicity to train, lack of specialized hardware requirements, and suitability for deployment on existing base-station processors and even mobile device processors in a future DSS system.  

\begin{figure}[t]
    \centering
    \includegraphics[width=0.9\linewidth]{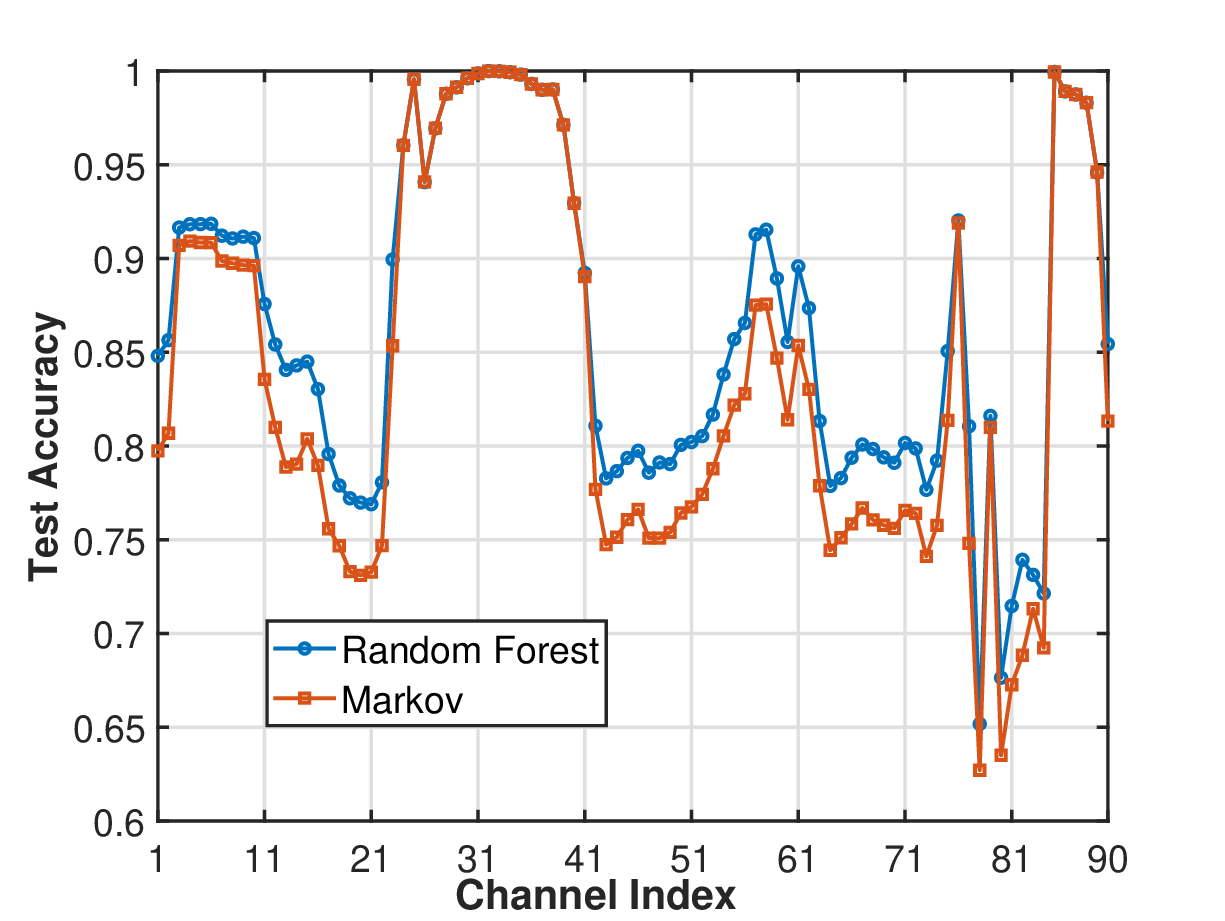}
    \caption{Test accuracy vs channels.}
    \label{fig:test_accuracy}
\end{figure}
\begin{figure}[t]
    \centering
    \includegraphics[width=0.9\linewidth]{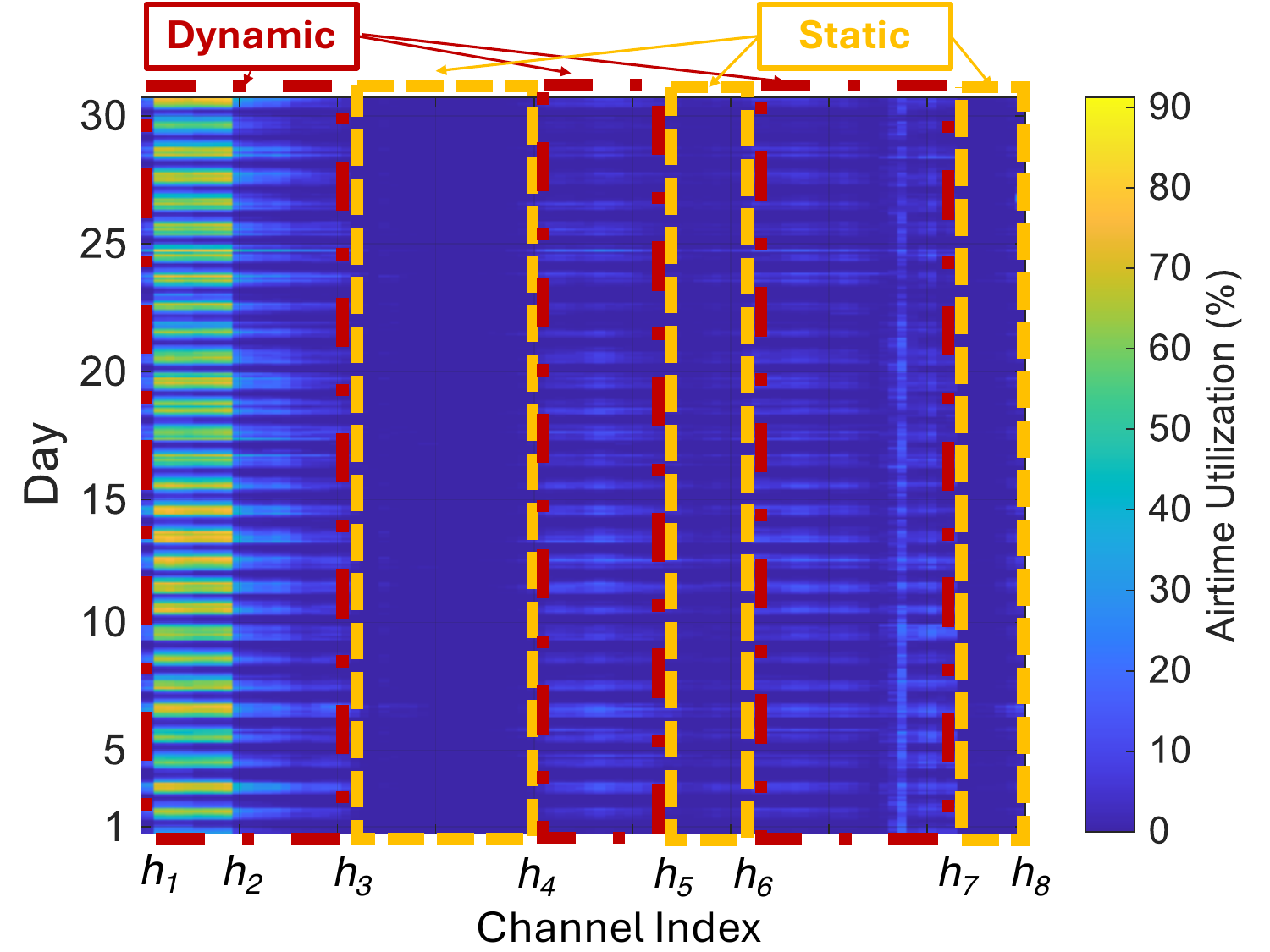}
    \caption{Dynamic and static channels.}
    \label{fig:dynamicChannel}
\end{figure}
Figure~\ref{fig:test_accuracy} compares the per-bin test accuracy of the Markov baseline with that of a representative machine-learning method, Random Forest.
It is observed that several frequency bins reach accuracy levels above 99\% for both methods; these bins correspond to non-dynamic channels whose occupancy remains almost always idle or occupied in the raw measurements, so any reasonable predictor achieves near-perfect accuracy.  
As shown in Fig.~\ref{fig:dynamicChannel}, airtime utilization, defined in~\cite{raouf2025spectrum} as the hourly spectrum usage percentage, remains nearly constant for non-dynamic (static) channels, which appear uniformly blue and exhibit minimal temporal variation, while dynamic channels show clear usage fluctuations over time.
In contrast, for bins with meaningful temporal variation, the Random Forest model consistently outperforms the Markov chain, often by a substantial margin.  
This behavior highlights the value of learning-based approaches when channel occupancy exhibits nontrivial dynamics and demonstrates the practical advantage of using a lightweight ML model, such as Random Forest, for short-term spectrum utilization forecasting.

\begin{figure}[t]
    \centering
    \includegraphics[width=0.8\linewidth]{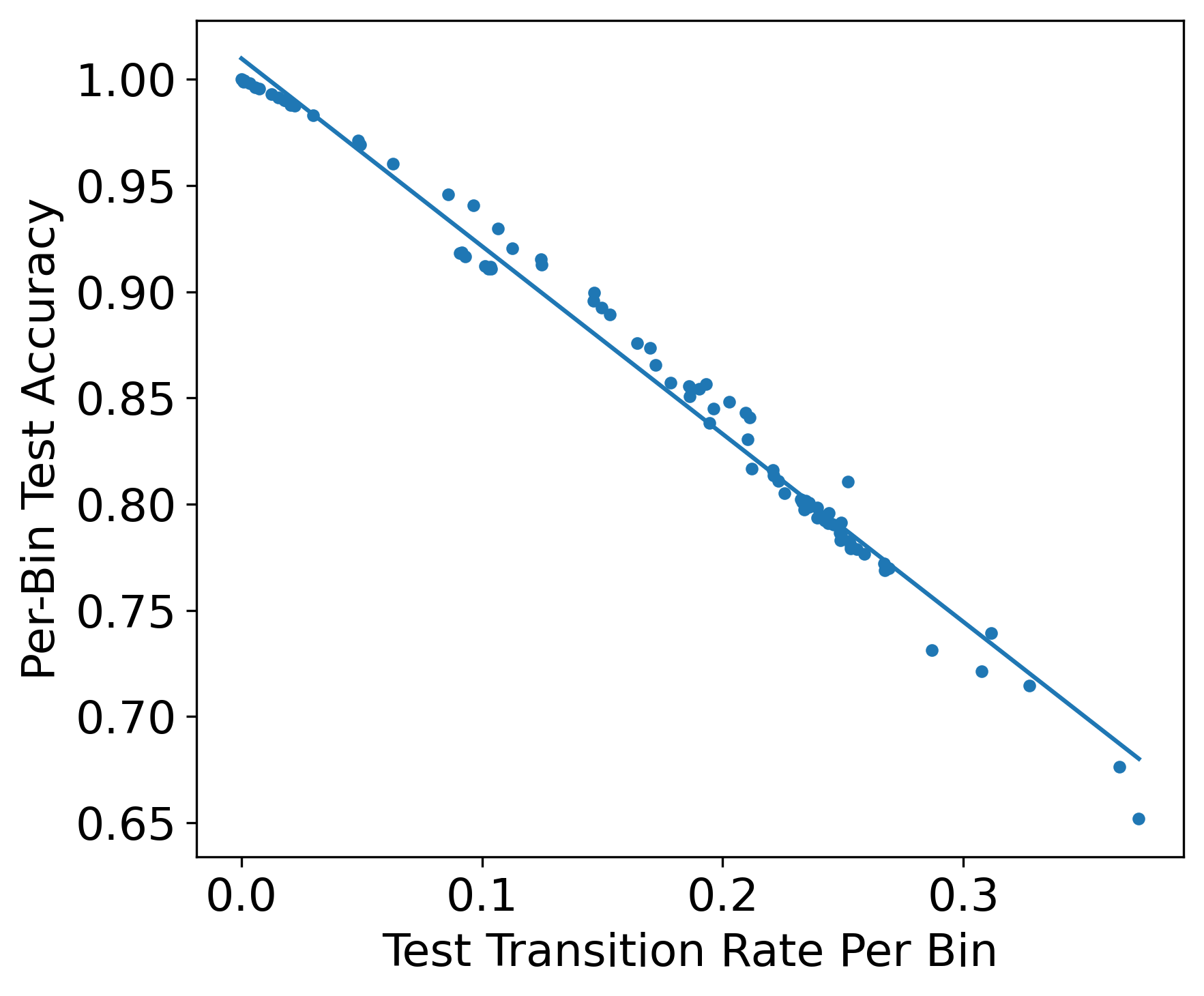}
    \caption{Accuracy vs channel dynamics.}
    \label{fig:acc_vs_dyn}
\end{figure}
Figure~\ref{fig:acc_vs_dyn} illustrates the relationship between the per-bin test accuracy of the Random Forest model and the corresponding transition rate, defined as the total number of state changes between idle and occupied states across all time transitions for each channel, which we use as a quantitative measure of channel dynamics.  
Each scatter point represents an individual frequency bin, while the superimposed line shows the least-squares linear fit.  
We observe a clear and nearly-linear trend: bins with low transition rates (i.e., almost static channels) achieve accuracy close to 1.0, whereas accuracy decreases steadily as the channel becomes more dynamic.  
This behavior is consistent with our earlier observations: when the occupancy state rarely changes, any prediction method performs well, but when the occupancy varies frequently, machine learning models such as Random Forest offer a significant advantage over the statistical baseline.  
The pronounced linear relationship in this figure further highlights how spectrum dynamics fundamentally influence prediction difficulty and demonstrates the ability of the Random Forest model to gracefully adapt to increasingly challenging channels.

\begin{figure}[t]
    \centering
    \includegraphics[width=0.9\linewidth]{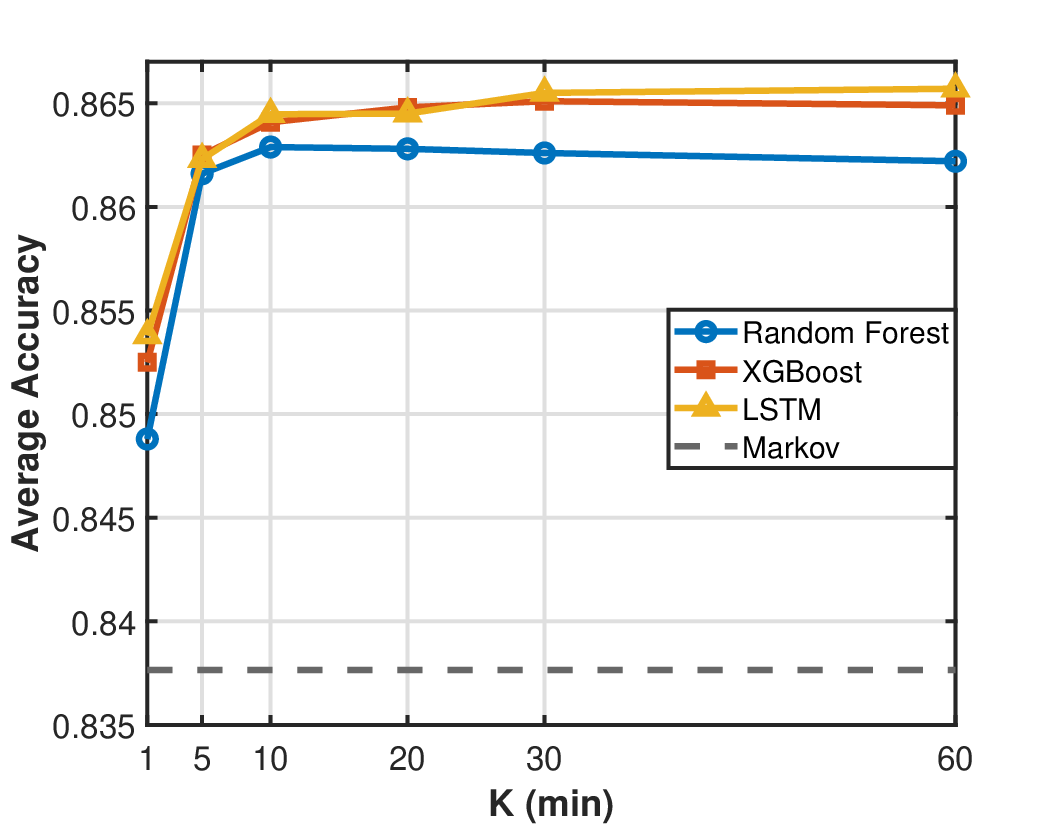}
    \caption{K sensitivity.}
    \label{fig:Ksense}
\end{figure}
Figure~\ref{fig:Ksense} examines the sensitivity of the learning methods to the history length~$K$, where a larger~$K$ incorporates more past observations in each training sample. 
As expected, the Markov baseline appears as a flat line because it does not use any history features.
For the three ML methods, the average accuracy initially improves when $K$ increases from very small values, since limited historical context cannot fully capture short-term temporal dependencies.  
However, when $K$ becomes excessively large, the performance begins to saturate or slightly degrade because too long histories introduce redundant or noisy information that does not contribute to the next-minute prediction.  
These results indicate that $K$ from 10 to 30, as a moderate amount of recent history, offers the best tradeoff, and that neither too little nor too much temporal context is beneficial for learning channel occupancy patterns.

\begin{figure}[t]
    \centering
    \includegraphics[width=1\linewidth]{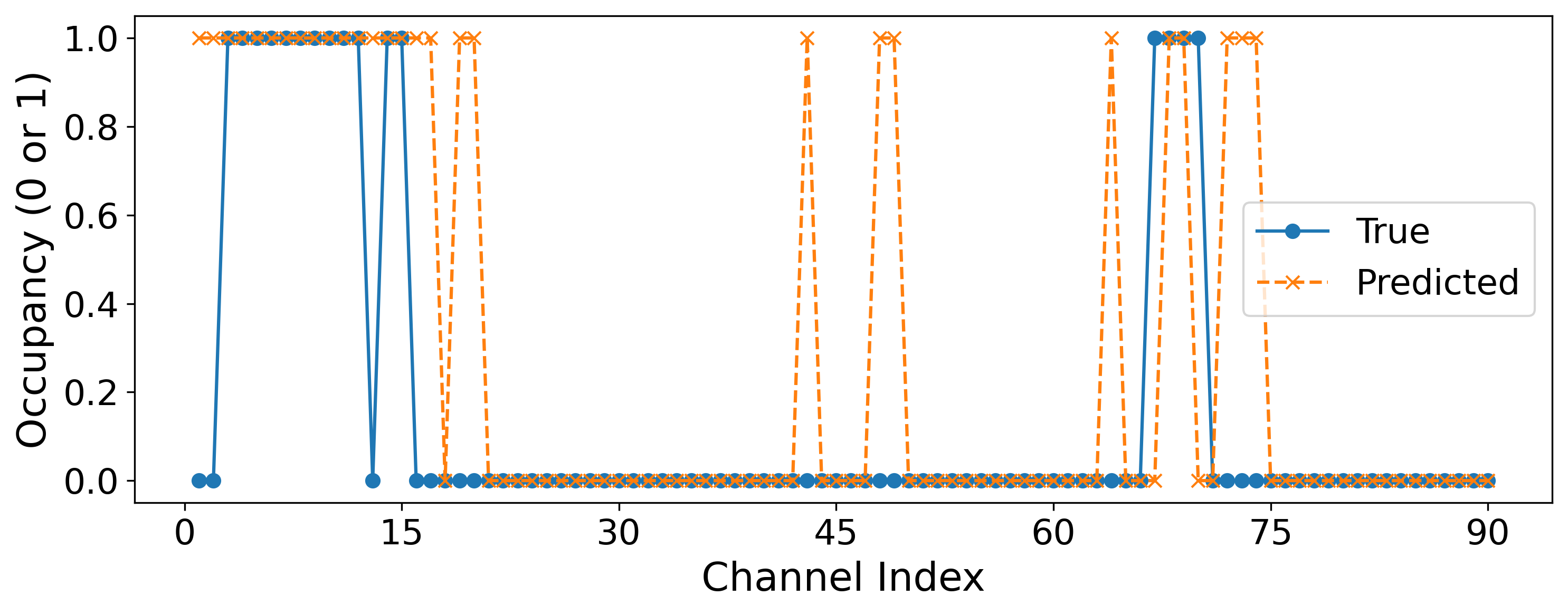}
    \caption{Example of Random Forest predictions at a single minute.}
    \label{fig:example_minute}
\end{figure}
Figure~\ref{fig:example_minute} provides an illustrative example of the predicted occupancy across all frequency bins for a single test minute, comparing the ground-truth pattern with the Random Forest predictions. 
Most errors visible in the figure occur near occupancy transitions: occasional false alarms, where idle bins are predicted as occupied, and missed detections, where short occupied intervals are predicted as idle. 
In contrast, bins exhibiting static behavior show almost no errors.
Although this is only one representative minute, it reflects the typical behavior observed in our experiments: the model performs almost perfectly on bins with no temporal variation and offers accurate short-term forecasts on bins exhibiting moderate dynamics, consistent with the aggregate results reported earlier.

\section{Conclusion} 
\label{sec: conclusion}
Channel availability prediction enables real-time and proactive dynamic spectrum allocation.
This study is based on 61 days of continuous wideband spectrum scanning measurement, from which channel occupancy is extracted through post-processing to construct a dataset suitable for supervised learning.
Four prediction approaches are evaluated, including a statistical first-order Markov chain and three machine learning methods—Random Forest, XGBoost, and LSTM. Their effectiveness in short-term channel occupancy prediction are accessed.
The first-order Markov model achieves an average accuracy of approximately 83\%, while the ML-based methods consistently exceed 86\% on average.
The results further indicate that prediction performance is highly dependent on the temporal dynamics of channel occupancy.
Data-driven learning methods offer clear advantages for short-term channel occupancy prediction in dynamic spectrum environments.
While the first-order Markov model performs adequately on static or low-variability channels, its performance degrades noticeably as temporal fluctuations increase.  
In contrast, Random Forest, XGBoost, and LSTM maintain consistently high performance across a wide range of dynamic regimes, even when trained on limited historical data. 
Sensitivity analysis reveals that a training history of 10 minutes offers the best performance-complexity tradeoff for Random Forest, whereas longer histories of approximately 30 minutes are optimal for XGBoost and LSTM.
Among these evaluated approaches, Random Forest emerges as the most practical option providing strong predictive performance while maintaining low computational complexity and ease of deployment, especially for on-device or edge deployments. 
A key insight is that increased model complexity does not necessarily translate to better short-horizon prediction performance on real spectrum traces.
These findings suggest that lightweight ML models, and Random Forest in particular, are well suited for integration into future dynamic spectrum sharing systems to enable accurate near-term spectrum utilization forecasting and more efficient dynamic spectrum access.
Future research directions include exploring additional machine learning models to further improve prediction accuracy, classifying frequency bins according to spectrum usage dynamics and user radio waveform characteristics, and extending spectrum utilization prediction to multi-sensor deployment distributed over a geographic area.

\bibliographystyle{IEEEtran}{}
\bibliography{refs}

@article{raouf2025spectrum,
  title={{Spectrum Usage Monitoring and Airtime Utilization: Insights from a Practical Case Study}},
  author={Raouf, Amir Hossein Fahim and Sun, Ruoyu and Poletti, Mark J},
  journal={IEEE Network},
  year={2025},
  publisher={IEEE}
}

@article{mark2024scte,
  title={{Spectrum Utilization: Nationwide
Measurements for New Spectrum Opportunities and Government Policy}},
  author={Poletti, Mark J and Sun, Ruoyu and Raouf, Amir Hossein Fahim},
  journal={SCTE TechExpo24},
  year={2024},
}

@article{rahilvtc,
  title={{An Automated Spectrum Utilization Monitoring Framework}},
  author={Gandotra, Rahil and Sun, Ruoyu and Poletti, Mark J and Mao, Jiayu and Guo, Hao},
  journal={submitted},
  year={2025},
}

@techreport{3gpp_tr_22_870,
  author      = "{3GPP}",
  title       = "{Study on 6G Use Cases and Service Requirements}",
  institution = "3rd Generation Partnership Project (3GPP)",
  type        = "TR",
  number      = "22.870",
  version     = "17.2.0",
  year        = "2025",
  month       = dec
}

@techreport{3gpp_s1_254270,
  author      = "{DISA, FirstNet, AT\&T, Cohere Technologies, CableLabs, and Erillisverkot}",
  title       = "{Use Case on Spectrum Scanning to Support Disaster Recovery}",
  institution = "3GPP SA1\#112 meeting",
  number      = "S1-254270",
  year        = "2025",
  month       = nov,
  address     = {Dallas, USA}
}

@techreport{fcc,
  author      = "{Federal Communications Commission}",
  title       = "{Radio Spectrum Allocation}",
  type        = "https://www.fcc.gov/engineering-technology/policy-and-rules-division/general/radio-spectrum-allocation",
}

@article{yin2012mining,
  title={{Mining Spectrum Usage Data: A Large-Scale Spectrum Measurement Study}},
  author={Yin, Sixing and Chen, Dawei and Zhang, Qian and Liu, Mingyan and Li, Shufang},
  journal={IEEE Transactions on Mobile Computing},
  volume={11},
  number={6},
  pages={1033--1046},
  year={2012},
  publisher={IEEE}
}

@article{wang2024deep,
  title={{Deep Learning Models for Spectrum Prediction: A Review}},
  author={Wang, Lei and Hu, Jun and Zhang, Chudi and Jiang, Rundong and Chen, Zengping},
  journal={IEEE Sensors Journal},
  volume={24},
  number={18},
  pages={28553--28575},
  year={2024},
  publisher={IEEE}
}

@article{chen2014predicting,
  title={{Predicting Spectrum Occupancies Using a Non-Stationary Hidden Markov Model}},
  author={Chen, Xianfu and Zhang, Honggang and MacKenzie, Allen B and Matinmikko, Marja},
  journal={IEEE wireless communications letters},
  volume={3},
  number={4},
  pages={333--336},
  year={2014},
  publisher={IEEE}
}

@inproceedings{wen2008autoregressive,
  title={{Autoregressive Spectrum Hole Prediction Model for Cognitive Radio Systems}},
  author={Wen, Zhigang and Luo, Tao and Xiang, Weidong and Majhi, Sudhan and Ma, Yunhong},
  booktitle={ICC Workshops-2008 IEEE International Conference on Communications Workshops},
  pages={154--157},
  year={2008},
  organization={IEEE}
}

@article{zhang2023gcrnn,
  title={{A-GCRNN: Attention Graph Convolution Recurrent Neural Network for Multi-Band Spectrum Prediction}},
  author={Zhang, Xile and Guo, Lantu and Ben, Cui and Peng, Yang and Wang, Yu and Shi, Shengnan and Lin, Yun and Gui, Guan},
  journal={IEEE Transactions on Vehicular Technology},
  volume={73},
  number={2},
  pages={2978--2982},
  year={2023},
  publisher={IEEE}
}

@article{li2025multi,
  title={{Multi-Dimensional Spectrum Prediction Using Closed-Form Continuous-Time Neural Network With Graph Attention}},
  author={Li, Ruicheng and Wang, Shufei and Huang, Hao and Yuen, Chau and Wang, Xianbin and Adachi, Fumiyuki and Gui, Guan},
  journal={IEEE Transactions on Vehicular Technology},
  year={2025},
  publisher={IEEE}
}

@inproceedings{alahakoon2025lstm,
  title={{LSTM-Based Spectrum Occupancy Prediction Performance with Thin Plate Spline Model}},
  author={Alahakoon, Dinushika Chathurangani and Kandeepan, Sithamparanathan and Yu, Xinghuo and Baldini, Gianmarco},
  booktitle={2025 International Conference on Information Networking (ICOIN)},
  pages={540--545},
  year={2025},
  organization={IEEE}
}

@article{xu2024spectrum,
  title={{Spectrum Prediction for Mobile Internet of Things Based on a DB-LSTM Algorithm}},
  author={Xu, Lingwei and Gao, Zhihe and Chen, Zhe and Wang, Jingjing and Fu, Yong and Li, Xingwang and Gulliver, T Aaron and Le, Khoa N},
  journal={IEEE Transactions on Vehicular Technology},
  volume={73},
  number={10},
  pages={15395--15406},
  year={2024},
  publisher={IEEE}
}

@article{mosavat2021prediction,
  title={{Prediction and Modeling of Spectrum Occupancy for Dynamic Spectrum Access Systems}},
  author={Mosavat-Jahromi, Hamed and Li, Yue and Cai, Lin and Pan, Jianping},
  journal={IEEE Transactions on Cognitive Communications and Networking},
  volume={7},
  number={3},
  pages={715--728},
  year={2021},
  publisher={IEEE}
}

@article{ji2025generative,
  title={{Generative Augmented Cascade Broad Learning for Lightweight Multi-Band Spectrum Prediction}},
  author={Ji, Niancong and Liu, Tiancheng and Zhang, Yibin and Wang, Qin and Ohtsuki, Tomoaki and Gui, Guan and Yuen, Chau and Adachi, Fumiyuki},
  journal={IEEE Transactions on Cognitive Communications and Networking},
  year={2025},
  publisher={IEEE}
}

@article{liu2025spectrumllm,
  title={{SpectrumLLM: Large Language Models for Next-Generation Spectrum Prediction}},
  author={Liu, Chao and Wang, Yu and Mao, Shiwen and Niyato, Dusit and Wang, Xianbin and Gui, Guan},
  journal={IEEE Wireless Communications},
  year={2025},
  publisher={IEEE}
}

@article{li2025novel,
  title={{A Novel Multi-Scale Time Fusion Transformer for Long-Range Spectrum Occupancy Prediction}},
  author={Li, Shuang and Sun, Yaxiu and Yue, Wenlu and Yao, Mengchen and Han, Yu and Gui, Guan and Lin, Yun and Xiang, Wei},
  journal={IEEE Transactions on Vehicular Technology},
  year={2025},
  publisher={IEEE}
}

@article{zhou2025spectrumfm,
  title={{SpectrumFM: A Foundation Model for Intelligent Spectrum Management}},
  author={Zhou, Fuhui and Liu, Chunyu and Zhang, Hao and Wu, Wei and Wu, Qihui and Quek, Tony QS and Chae, Chan-Byoung},
  journal={arXiv preprint arXiv:2505.06256},
  year={2025}
}

@inproceedings{sagduyu2024joint,
  title={{Joint Sensing and Task-Oriented Communications with Image and Wireless Data Modalities for Dynamic Spectrum Access}},
  author={Sagduyu, Yalin E and Erpek, Tugba and Yener, Aylin and Ulukus, Sennur},
  booktitle={2024 IEEE International Symposium on Dynamic Spectrum Access Networks (DySPAN)},
  pages={57--62}
}

@article{xiong2013spectrum,
  title={{Spectrum Sensing in Cognitive Radio Networks: Performance Evaluation and Optimization}},
  author={Xiong, Gang and Kishore, Shalinee and Yener, Aylin},
  journal={Physical Communication},
  volume={9},
  pages={171--183},
  year={2013},
  publisher={Elsevier}
}

@article{guovtc,
  title={{LarS-Net: A Large-Scale Framework for Network-Level Spectrum Sensing}},
  author={Guo, Hao and Sun, Ruoyu and Raouf, Amir Hossein Fahim and Gandotra, Rahil and Mao, Jiayu and Poletti, Mark J},
  journal={submitted},
  year={2025},
}

@ARTICLE{lstm,
  author={Hochreiter, Sepp and Schmidhuber, Jürgen},
  journal={Neural Computation}, 
  title={{Long Short-Term Memory}}, 
  year={1997},
  volume={9},
  number={8},
  pages={1735-1780},
  doi={10.1162/neco.1997.9.8.1735}}

\end{document}